\documentclass{ws-mpla}
\usepackage[super]{cite}
\usepackage{graphicx}
\usepackage{hyperref}
\usepackage[usenames,dvipsnames]{color}
\usepackage{graphicx}
\usepackage{amsmath}
\usepackage{amsfonts}
\usepackage{dsfont}
\usepackage{dcolumn}
\usepackage{bm}
\usepackage{slashed}
\usepackage{pstricks}
\usepackage{slashed}
\usepackage{multirow}
\usepackage{array}
\usepackage{rotating}
\usepackage{xcolor}
\usepackage{epstopdf}
\usepackage{fancybox}
\usepackage{fancyvrb}
\usepackage[normalem]{ulem}
\usepackage{color}
\usepackage{mathdots}
\usepackage{mathrsfs}
\usepackage{multirow}
\usepackage{esint}

\def \d{{\rm d}}

\usepackage{cleveref}
\crefname{equation}{Eq.}{Eqs.}
\crefname{figure}{Fig.}{Figs.}
\crefname{table}{Table}{Tables}
\crefname{Section}{Section}{Sections}
\allowdisplaybreaks

\begin{document}

\markboth{Wan-Zhe Feng and Pran Nath}
{Baryogenesis and Dark Matter in $U(1)$ Extensions}

%%%%%%%%%%%%%%%%%%%%% Publisher's Area please ignore %%%%%%%%%%%%%%
%\catchline{}{}{}{}{}
%%%%%%%%%%%%%%%%%%%%%%%%%%%%%%%%%%%%%%%%%%%%%%%%%%%%%%%%%%%%%%

\title{Baryogenesis and Dark Matter in $U(1)$ Extensions}

\author{Wan-Zhe Feng, Pran Nath}

\address{Department of Physics, Northeastern University, Boston, MA 02115-5000, USA\\
w.feng@northeastern.edu, p.nath@neu.edu}

\maketitle

%\pub{Received (Day Month Year)}{Revised (Day Month Year)}

\begin{abstract}
A brief review is given of some recent works where baryogenesis and dark matter
have a common origin within the $U(1)$ extensions of the standard
model and of the minimal supersymmetric standard model.
The models considered generate the desired baryon asymmetry and the dark matter to baryon ratio.
In one model all of the fundamental interactions do not violate lepton number,
and the total $B-L$ in the Universe vanishes.
In addition, one may also generate a normal hierarchy of neutrino masses and mixings in conformity with the current data.
Specifically one can accommodate $\theta_{13}\sim 9^{\circ}$ consistent with the data from Daya Bay reactor neutrino experiment.

\keywords{Baryogenesis, Dark Matter, Neutrino, Daya Bay}
\end{abstract}

%\ccode{PACS Nos.: 95.35.+d,  12.60.Jv.}

\section{Introduction}	

Three of the important puzzles in cosmology relate to the origin of baryon asymmetry in the Universe,
the nature of dark matter and the cosmic coincidence that the amount of dark matter and visible matter are comparable.
%i.e.,  one has~\cite{Ade:2013sjv} ${\Omega_{\rm DM} h_0^2}\big/{ \Omega_{\rm B} h_0^2}\approx 5.5$.
The fact that dark matter and visible matter are comparable in size points to the possibility of a common origin of the two.
Here we discuss classes of models where baryon asymmetry and dark matter have a common origin
within the framework of $U(1)$ extensions of the standard model (SM) and of the minimal supersymmetric standard model (MSSM)~\cite{Feng:2013wn,Feng:2013zda,Feng:2012jn}.

The basic tenets of generating matter over anti-matter are well-known
and consist of three conditions~\cite{Sakharov:1967dj}:
the existence of baryon (or lepton) number violation, the presence of C and CP violating interactions, and out of equilibrium processes.
One suggestion for explaining the comparable size of dark matter and visible matter is the
so-called asymmetric dark matter hypothesis~\cite{Kaplan:2009ag}
where the dark particles are in thermal equilibrium with the SM (MSSM) particles in the early universe,
and thus their chemical potentials are of the same order.
The satisfaction of dark matter and visible matter ratio
${\Omega_{\rm DM} }/{ \Omega_{\rm B}}\approx 5.5$~\cite{Ade:2013sjv} can then be achieved
via a constraint on the dark matter mass (for reviews see~\cite{review}).
More specifically, the asymmetry can transfer from the visible sector to the dark sector via the asymmetry transfer interaction
$\mathcal{L}_{\rm asy} = \frac{1}{M^n_{\rm asy}}\mathcal{O}_{\rm DM} \mathcal{O}_{\rm asy}$~\cite{Kaplan:2009ag}, where $M_{\rm asy}$ is the scale of the interaction,
$\mathcal{O}_{{\rm asy}}$ is an operator constructed
from SM (MSSM) fields which carries a non-vanishing $B-L$ quantum number
while $\mathcal{O}_{{\rm DM}}$ carries the opposite $B-L$ quantum number.
This interaction would decouple at some temperature
greater than the dark matter mass.  As the Universe cools down,
the dark matter asymmetry freezes at the order of the baryon asymmetry,
which explains the observed relation between the amount of baryon and dark matter.
In~\cite{Feng:2012jn} we discussed asymmetric dark matter in the $U(1)_{L_\mu - L_\tau}$ and $U(1)_{B-L}$
Stueckelberg extensions of the SM and of MSSM~\cite{Stueckelberg}.
%In these works, the pre-existing baryon asymmetry is assumed.

In what follows we discuss two model classes where baryon asymmetry and dark matter have a common origin
(for related works see~\cite{DtoV,CoG}).
For the first model class,
dark matter is generated via the decay of some primordial fields and
the asymmetry created by the CP violating decays is
then transferred to the visible sector via the asymmetry transfer interaction~\cite{Feng:2013wn}.
In the second model class,
leptogenesis takes place with all the fundamental interactions conserving lepton number
and leptogenesis consists in generating equal and opposite lepton numbers in the visible and dark sectors~\cite{Feng:2013zda}.
Subsequently the sphaleron processes transmute a part of the lepton asymmetry into baryon asymmetry.
In this model class the total $B-L$ number in the Universe is exactly conserved.
%These ideas have recently been pursued by several authors~\cite{Shelton:2010ta,Davoudiasl:2010am}.

In the model classes referred to above the stability of dark matter is protected by the $U(1)$ gauge symmetry.
A kinetic mixing between the $U(1)$ and $U(1)_Y$ gauge bosons
allows for dissipation of the symmetric component of dark matter through the exchange of  the $U(1)$ gauge boson.
An alternative way of depleting the symmetric component of  dark matter is assuming that the $U(1)$ gauge boson is massless (dark photon).
Majorana mass terms for dark particles are forbidden.
Consequently, the dark matter asymmetry generated in the early universe would not be washed out by oscillations.

\section{Baryogenesis from Dark Sector}

We first discuss the model class where primordial fields decay
into dark matter and create an asymmetry.
The dark matter asymmetry then transmutes into lepton and baryon asymmetries.

\subsection{The model}

Here we work in a supersymmetric framework.\footnote{
In the non-supersymmetric case, the simplest model with interaction $\mathcal{L} \sim \lambda_i  N_i \bar X^c X' + h.c.$,
where $\lambda_i$ are complex coupling constants, $N_i$ are complex scalars and $i\geq2$, $X,X'$ are Dirac fermions, does not work.
Namely, the decay of $N_i$ as well as $N_i^*$ does not generate an asymmetry between $X,X'$ and $\bar{X},\bar{X}'$.}
We assume that in the early universe there exist several $\hat{N}_i$ fields ($i \geq 2$) with masses $M_i$,
where $\hat N=(N, \tilde{N})$ and $N$ is the Majorana field and $\tilde N$ is the super-partner field.
The scalar field of the lightest $\hat{N}_i$ superfields could play the role of the inflaton,
and $\hat{N}_i$ can also be right-handed neutrinos as suggested in earlier works.
The dark sector is comprised of $(\hat X, \hat{X}^c, \hat X', \hat{X}'^c)$ which are charged under the gauge group $U(1)_x$ with charges $(+1,-1,-1,+1)$
while the MSSM fields are not charged under $U(1)_x$.
We assume the $\hat{N}_i$ carry a non-vanishing lepton number $+2$,
$\hat X, \hat X'$ carry lepton number $-1$ and $\hat{X}^c, \hat{X}'^c$ carry lepton number $+1$.
The superpotential of the model is given by
\begin{equation}
W=
\lambda_i \hat{N}_i \hat{X} \hat{X}' +  \frac{1}{M_{\rm asy}^2} \hat{X}\hat{X}' (L H_u)^2
+ m \hat X \hat{X}^c + m' \hat X' \hat{X}'^c\,,
\label{Wfull}
\end{equation}
where the couplings $\lambda_i$ are assumed to be complex.
$W$ is invariant under both $U(1)_x$ and lepton number,
and the first term is responsible for generating an asymmetry in the dark sector
whereas the second term is responsible for transferring the asymmetry generated in the dark sector to the visible sector.
Finally we add mass terms for $\hat{N}_i$ to the superpotential,
i.e.,  a term $W \sim \frac{1}{2} M_i \hat{N}_i \hat{N}_i$,  which violates lepton number.

We assume the mass hierarchy $M_i \gg  m+ m'$ so that in the early universe,
and the out-of-equilibrium decays of $\hat{N}_i$ generates dark matter through
$N_i \to X \tilde X', \tilde X X', \bar{X} \tilde X'^*, \tilde X^* \bar{X}'$
and $\tilde{N}_i \to  X X',   \bar X \bar X'$.
Further, the CP violation due to the complex couplings  $\lambda_i$ generates
an excess of $X,X'$ over their anti-particles $\bar X, \bar X'$ carrying the opposite lepton numbers.
Thus the decays of $N_i$ produce a lepton number asymmetry in the dark sector.
The lepton asymmetry generated in this fashion in the dark sector is
then transferred to the visible sector through the asymmetry transfer interaction, and thus leptogenesis occurs.
Finally, a part of lepton number asymmetry  of the visible sector then transmutes to baryon number asymmetry via
the sphaleron interactions.
In the simplest model we have $i=2$,
and we assume $\hat{N}_2$ mass $M_2$ is much larger than $\hat{N}_1$ mass $M_1$.

\begin{figure}[t!]
\begin{center}
\includegraphics[scale=0.72]{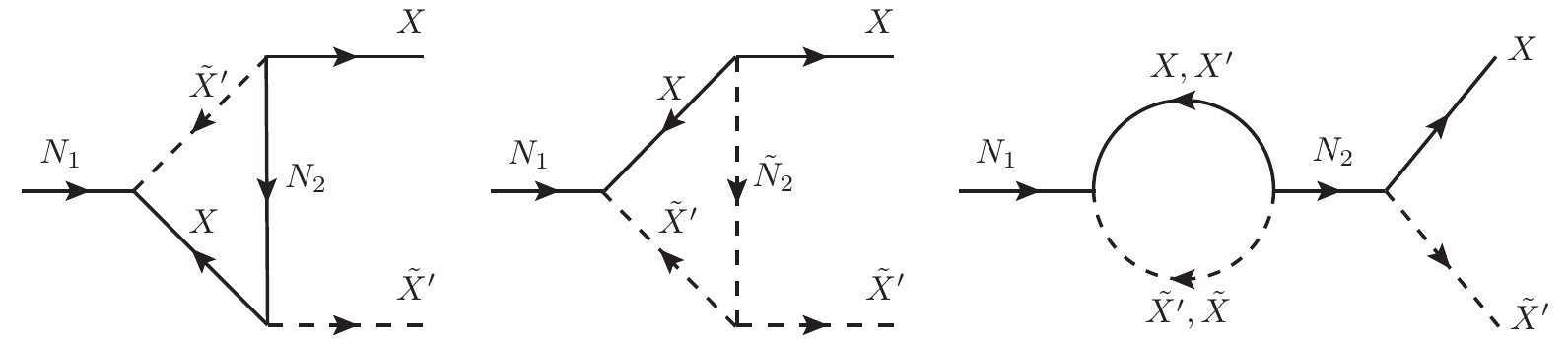}
\vspace*{8pt}
 \caption{Loop diagrams responsible for the
genesis of dark matter asymmetry from the decay of $N_1$ to final states $X\tilde X'$
and there are similar diagrams for the decay of the $N_1$ to the final states $\tilde X X'$,
and for the decay of $\tilde N_1$ to $X X'$ and to $\tilde{X} \tilde{X}'$.}
\label{DAsy}
\end{center}
\end{figure}

The dark matter asymmetry arises from the interference of the
one-loop diagrams shown in Fig.~\ref{DAsy} with the tree-level diagrams,
similar to the conventional leptogenesis diagrams~\cite{Early}.
The asymmetries, i.e.,  the excess of
$\hat{X},\hat{X}'$ over their anti-particles $\overline{\hat{X}},\overline{\hat{X}'}$
are measured by $\epsilon_{X\tilde{X}'}, \epsilon_{\tilde{X}X'}, \epsilon_{XX'}, \epsilon_{\tilde{X} \tilde{X}'}$~\cite{Feng:2013wn}
where the lower indices of $\epsilon$ denote the final state particles.
There are two types of loops involved: vertex contribution and wave contribution as shown in Fig~\ref{DAsy}.
It's straight forward to compute the above asymmetry parameters $\epsilon_{X\tilde{X}'}$ etc.
It turns out that the contributions of the vertex diagrams and the wave diagrams satisfy the following relations
\begin{gather}
\epsilon_{X\tilde{X}'}^{vertex}=\epsilon_{\tilde{X}X'}^{vertex}=\epsilon_{XX'}^{vertex}=\epsilon_{\tilde{X}\tilde{X}'}^{vertex}\equiv\epsilon^{vertex}\,,\\
\epsilon_{X\tilde{X}'}^{wave}=\epsilon_{\tilde{X}X'}^{wave}=\epsilon_{XX'}^{wave}=\epsilon_{\tilde{X}\tilde{X}'}^{wave}\equiv\epsilon^{wave}\,.
\end{gather}
Specifically, we have
\begin{align}
\label{vw1}
\epsilon^{vertex} & =-\frac{1}{8\pi}\frac{{\rm Im}(\lambda_{1}^{2}\lambda_{2}^{*2})}{|\lambda_{1}|^{2}}\frac{M_{2}}{M_{1}}\ln\frac{M_{1}^{2}+M_{2}^{2}}{M_{2}^{2}}\,,\\
\epsilon^{wave} & =-\frac{1}{8\pi}\frac{{\rm Im}(\lambda_{1}^{2}\lambda_{2}^{*2})}{|\lambda_{1}|^{2}}\frac{M_{1}(M_{1}+M_{2})}{M_{2}^{2}-M_{1}^{2}}\,.
\label{vw2}
\end{align}
Thus the total asymmetry parameter is the sum of the vertex and the wave contributions
and in the limit $M_{2}\gg M_{1}$, we obtain
\begin{equation}
\epsilon=\epsilon^{vertex}+\epsilon^{wave}\approx-\frac{1}{4\pi}\frac{{\rm Im}(\lambda_{1}^{2}\lambda_{2}^{*2})}{|\lambda_{1}|^{2}}\frac{M_{1}}{M_{2}}\,.
\label{epsi}
\end{equation}

The total excess of $X,\tilde{X},X',\tilde{X'}$ over $\bar{X},\tilde{X}^{*},\bar{X}',\tilde{X'}^{*}$
generated by the decay of $\hat{N}_1$ is given by $\Delta n_X \approx 2\kappa s \epsilon \big/ g_*$,
where $s$ is the entropy, $g_* \approx 228.75$ is the entropy degrees of freedom for MSSM,
and $\kappa$ is a washout factor due to inverse processes
$X+\tilde X', \tilde X+ X' \to N$ and $X+ X', \tilde X+ \tilde X' \to \tilde N$ and in our analysis we set $\kappa =0.1$.
The excess of $\hat{X}, \hat{X}'$ then give rise to a non-vanishing $(B-L)$-number in the early universe:
$(B-L)_{\rm t} = (+1)\times \Delta n_X \approx 2\kappa s \epsilon \big/ g_*$, where $(B-L)_{\rm t}$ is the total $B-L$ in the Universe
and $+1$ indicates each of $X, X'$ carries a $B-L$ number $+1$.

The $B-L$ asymmetry generated in the visible sector through the asymmetry transfer interaction
can be obtained by using the standard thermal equilibrium method introduced in~\cite{Harvey:1990qw}.
For very high temperatures the MSSM fields are ultra-relativistic, hence
MSSM fields and dark particles are in thermal equilibrium,
which gives rise to relations among
their chemical potentials~\cite{Harvey:1990qw,Feng:2012jn}.
These relations  allow us to express the chemical potentials of all the MSSM fields
in terms of the chemical potential of one single field, e.g., $\mu_{L}$,  the chemical potential of the left-handed lepton doublet.
Similarly other quantities of interest, i.e., the total lepton number $L$, the total baryon number $B$, and
the net $B-L$ in the visible sector can all be expressed in terms of $\mu_{L}$.
Specifically we have $(B-L)_{\rm v}  =-\tfrac{237}{7}\mu_L$, where $(B-L)_{\rm v}$ is the $B-L$ in the visible sector.
Here we assume the asymmetry transfer interaction would decouple above the supersymmetry breaking scale,
thus the asymmetry would transfer from the dark sector to the visible sector when all of the MSSM particles are active in the thermal bath.
Hence dark particles are in thermal equilibrium with all of the MSSM particles, which gives
$\mu_{\hat{X}}+\mu_{\hat{X}'}= -\mu_{\hat{X}^c} -\mu_{\hat{X}'^c} = -\frac{22}{7}\mu_{L}$.
Thus the total dark particle number is given by $X = \frac{44}{79} (B-L)_{\rm v}$.

The dark matter mass is determined using the constraint
$\Omega_{\rm DM} / \Omega_{\rm B} = (X \,m_{\rm DM}) / (B \,m_{\rm B}) \approx 5.5$,
where  $m_{\rm DM}$ is the mass of the dark matter particle
and $m_B$ is the baryon mass which is taken to be $m_{\rm B}\sim 1$ GeV.
An important subtlety here is that although the total dark particle number is fixed after the
asymmetry transfer interaction decouples,
the total baryon number changes after this decoupling
because of the sphaleron processes.
As explained in detail in~\cite{Feng:2012jn},
the total baryon number to be used in the computation of $\Omega_{\rm DM} / \Omega_{\rm B}$
is $B_{\rm final}$ after the sphaleron processes decouple.
Thus one has $m_{\rm DM} =  (B_{\rm final} / X) \cdot 5.5~{\rm GeV}$
where $B_{\rm final} = \frac{30}{97} (B-L)_{\rm v} \approx 0.31 (B-L)_{\rm v}$~\cite{Feng:2012jn}.
This leads to  $m_{\rm DM} \approx 3.01~{\rm GeV}$.
The astrophysical constraint $B_{\rm final} / s \sim  6  \times 10^{-10}$~\cite{Beringer:1900zz},
can be satisfied with $\epsilon \sim 4 \times 10^{-6}$,
which sets bounds for the complex couplings $\lambda_i$ and the ratio $M_1/M_2$.

\subsection{Physics of the dark sector}

In order to achieve a viable model one needs to dissipate the symmetric component of
dark matter. This can be achieved  by gauge kinetic energy mixing
of $U(1)_x$ and $U(1)_Y$~\cite{Holdom:1985ag}.
The thermally produced dark matter  and its anti-matter can annihilate efficiently into SM particles
through the $Z'$ boson exchange with a Breit-Wigner enhancement~\cite{DMRD,NathWig,Celis:2016ayl}.
The kinetic mixing does not generate couplings between the photon and dark sector particles and thus
dark matter carries no milli-charge.
Consequently there are no experimental constraints  from the limits on milli-charges on the parameter $\delta$
which enters in the gauge kinetic energy mixing of $U(1)_x$ and $U(1)_Y$.
Thus the strongest experimental constraints on the $Z'$ boson mass and its coupling to the visible sector
come from corrections to $g_{\mu}-2$ as well as LEP II constraints.
These lead to the limit $\delta \lesssim 0.001$.
With such constraints, one can  deplete the symmetric component of dark matter
in sufficient amounts, i.e., less than $10\%$ of the total dark matter relic abundance.

An alternative way of depleting the symmetric component of  dark matter is
assuming that the $U(1)_x$ gauge boson is massless (dark photon).
Then the symmetric component of the dark matter could  annihilate
into the $U(1)_x$ dark photons and become radiation in the early universe.
As shown in~\cite{Blennow:2012de},
the constraints on the number of extra effective neutrino species $\Delta N_{\rm eff}$,
can be satisfied for a large class of asymmetric dark matter models.

Such dark matter can scatter from quarks within a nucleon through the t-channel exchange of the $Z'$ boson.
The spin-independent dark matter-nucleon cross
section can be approximately written as
$\sigma_{{\rm SI}}\sim 4\delta^{2} g_{x}^{2} g_{Y}^{2} \cos^{4}\theta_{W}\mu_{n}^{2} \big/ \pi m_{Z'}^{4}$,
where $\mu_{n}$ is the dark matter-nucleon reduced mass.
For our model we find $\sigma_{{\rm SI}}\sim10^{-37}~{\rm cm}^{2}$,
which is just on the edge of sensitivity of the CRESST~I experiment~\cite{Angloher:2002in}.
Thus improved experiment in the future in the low dark matter mass region
with better sensitivities should be able to test the model.

As in the supersymmetric case,
the $U(1)_x$ gaugino $\chi$  is given a soft mass $\mathcal{L}_{\chi} = m_{\chi} \bar \chi \chi$.
It can then decay into $X\tilde X$ or $X' \tilde X'$ via the supersymmetric interaction
$\mathcal{L} \sim \chi X \tilde X+ \chi X' \tilde X' + h.c.$,
where we  assume  $m_{\chi} > m_X + m_{\tilde X}$.
Thus the gaugino $\chi$ decays into dark particles and is removed from the
low energy spectrum.
One important aspect of the supersymmetric case is that it presents a multi-component picture of dark matter.
The total dark matter relic abundance consists of dark sector particles $(\hat X, \hat{X}^c, \hat X', \hat{X}'^c)$
as well as the conventional  lightest supersymmetric particle with R-parity, i.e., the (lightest) neutralino.
There exists a significant part of the parameter space of MSSM where the relic density of neutralinos can be  10\% or less of the current relic density~\cite{Feng:2012jn}.
The analysis of~\cite{Feng:2012jn} shows that even with 10\% of the relic density,
the neutralino dark matter would be still accessible in dark matter searches.
Thus this feature also offers a direct test of the model in neutralino dark matter searches.
However, the leptonic dark matter would be difficult
to see in direct searches for dark matter as well as in collider experiments because of its small couplings
to the visible sector via the $Z'$ boson exchange.
Future colliders with higher sensitivity and accuracy may have
the possibility to explore the $Z'$ boson with tiny couplings to the SM particles.

\section{Cogenesis of Baryon Asymmetry and Dark Matter}

We discuss now a model class~\cite{Feng:2013zda} where
leptogenesis takes place with all fundamental interactions not violating lepton number,
and the total $B-L$ number in the Universe vanishes.
Such leptogenesis leads to equal and opposite lepton numbers in
the visible sector and the dark sector.
Part of the lepton number generated in the visible sector
subsequently transfers to the baryonic sector via sphaleron interactions.

\subsection{The model}

We begin by considering the set of fields  $N_i, \psi, \phi, X, X'$
with lepton number assignments $(0, +1, -1, +1/2, +1/2)$.
Here $N_i$ ($i \geq 2$) are Majorana fermions,
$\psi, X,X'$ are Dirac fields and $\phi$ is a complex scalar field.
The fields $N_i, \psi, \phi$ are heavy and will decay into lighter fields
and eventually disappear.
The dark sector is constituted of two fermionic fields $X,X'$, which
as indicated above each carry a lepton number $+1/2$ and
are oppositely charged under the dark sector gauge group $U(1)_x$ with gauge charges $(+1,-1)$.
All other fields are neutral under $U(1)_x$.  We assume their interactions to have the following form
which conserve both the lepton number and the $U(1)_x$ gauge symmetry
\begin{equation}
\mathcal{L} =\lambda_i \bar N_i \psi \phi + \beta\, \bar \psi L H
+ \gamma\, \phi \bar X^c X' + h.c.\,,
\label{1.1}
\end{equation}
where the couplings $\lambda_i$ are assumed to be complex and the couplings $\beta,\gamma$ are
assumed to be real. In addition we add mass terms so that
\begin{equation}
- \mathcal{L}_m= M_i \bar N_i N_i +  m_1 \bar \psi \psi + m_2^2 \phi^* \phi
  +m_X \bar X X + m_{X'} \bar X' X' \,.
\label{1.2}
\end{equation}
Here $N_i$ have Majorana masses, while $\psi, X, X'$ have Dirac masses.
We assume the mass hierarchy $M_i \gg  m_1 + m_2$, $m_1\sim m_2 \gg m_X+ m_{X'}$.
%We will see soon that $m_X,m_{X'}$ are around 1~GeV.
Consistent with the above constraint,
$m_1, m_2$, the masses of $\psi$ and $\phi$,
could span a wide range from TeV scale to scales much higher.

In the early universe, the out-of-equilibrium decays of the heavy Majorana fields $N_i$
produce a heavy Dirac field $\psi$ and a heavy complex scalar field $\phi$.
The CP violation due to the complex couplings  $\lambda_i$ generates
an excess of $\psi,\phi$ over their anti-particles $\bar{\psi},\phi^*$
which carry the opposite lepton numbers.
Since the lepton number carried by $\psi$ and $\phi$ always sums up to zero,
the out-of-equilibrium decays of $N_i$ do not generate an excess of lepton number in the Universe.
Further, $\psi$ and $\phi$ (as well as their anti-particles) produced in the decay of the Majorana fields $N_i$
will sequentially decay, with $\psi$ (and its anti-particle) decaying into the visible sector fields and $\phi$
(and its anti-particle) decaying into
the dark sector fields.  Their decays thus produce a net lepton asymmetry in the visible sector and
a lepton asymmetry of opposite sign in the dark sector.
We note that the absence of the decays
$\psi \to \bar{X}+ X'$ and $\phi^* \to L+H$ guarantees
that leptonic asymmetries of equal and opposite sign are generated
in the visible and in the dark sectors.
Indeed, right after the heavy Majorana fermions $N_i$ have decayed completely,
and created the excess of $\psi,\phi$ over $\bar{\psi},\phi^*$,
equal and opposite lepton numbers are already assigned to
the visible sector and the dark sector.
It is clear from the above analysis that there is no violation of lepton number in the entire process of
generating the leptonic asymmetries.
We further note that while sphaleron interactions are active
during the period when the leptogenesis and the genesis of (asymmetric) dark matter occur,
they are not responsible for creating a net $B-L$ number in the visible sector,
though they do play a role in transmuting a part of the lepton number into baryon number in the visible sector.

One can estimate on general grounds  the mass of the dark particles in this model for the cosmic coincidence to occur.
Since the total $B-L$ in the Universe vanishes,
the  $B-L$ number in the visible sector is equal in magnitude and opposite in sign
to the lepton number created in the visible sector right after $N_i$ have completely decayed
(the decay of $N_i$ does not generate any baryon asymmetry),
and thus is equal to the lepton number in the dark sector, i.e., $   (B-L)_{\rm v} = L_{\rm d}$
where the indices ${\rm v,d}$ denote the visible sector and the dark sector respectively.
We are interested in the relative density of particle species at the time when
the sphaleron interactions go out of the thermal equilibrium.
After the decoupling of the sphaleron interactions $B$ and $L$ are separately conserved
and correspond to the $B$ and $L$ seen today.
Recall the final (currently observed) value of the baryon number density $B_{\rm final}\approx 0.31 (B-L)_{\rm v}$~\cite{Feng:2012jn},
assuming that $X$ and $X'$ have the same mass,
we obtain $m_X = m_{X'} \approx 0.85~{\rm GeV}$.

\begin{figure}[t!]
\begin{center}
\includegraphics[scale=0.68]{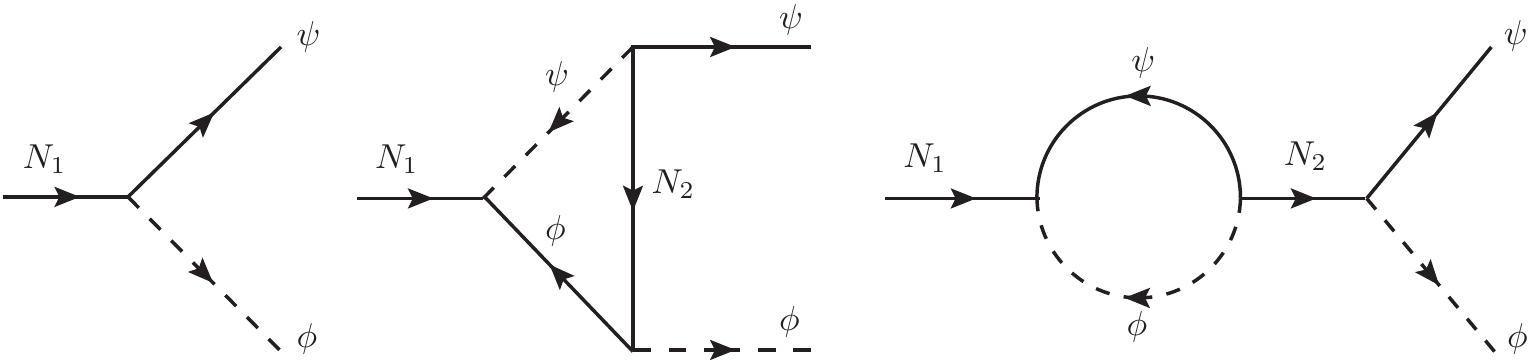}
\vspace*{8pt}
 \caption{Generation of asymmetry in $\psi,\phi$ over their antiparticles $\bar\psi,\phi^*$ from the decay of the Majorana field $N_1$. The lepton number is conserved in these processes.}
\label{Cog}
\end{center}
\end{figure}

We turn now to the detail of the generation of
the asymmetry between $\psi,\phi$ and $\bar{\psi}, \phi^*$.
We assume there are two Majorana fields $N_1$ and $N_2$
with $N_2$ mass $M_2$ being much larger than the $N_1$ mass $M_1$, i.e., $M_2 \gg M_1$.
The diagrams that contribute to it are shown
in~\cref{Cog} where the Majorana particles  $N_i$ decay into the Dirac fermion $\psi$
and the complex scalar $\phi$ with $\psi$ and $\phi$
carrying opposite lepton numbers while the Majorana fields $N_i$ carry no lepton number.
%As is well-known one needs an interference of the tree and the loop diagrams to create the asymmetry.
%The loop diagrams consist of  a vertex diagram and  a wave function diagram
%as shown in~\cref{fig2}.
In this case the asymmetry arising from the excess of  $\psi,\phi$ over $\bar{\psi}, \phi^*$ is given by
\begin{equation}
\epsilon
 =\frac{\Gamma(N_{1}\to \psi\phi)-\Gamma(N_{1}\to\bar\psi\phi^*)}
{\Gamma(N_{1}\to \psi\phi)+\Gamma(N_{1}\to\bar\psi\phi^*)} \simeq
-\frac{1}{8\pi}\frac{{\rm Im}(\lambda_{1}^{2}\lambda_{2}^{*2})}{|\lambda_{1}|^{2}}\frac{M_{1}}{M_{2}}
\,,
\label{ASM}
\end{equation}
where we have included both the vertex contribution and the wave contribution.
Since the dark sector does not communicate with the visible sector,
$(B-L)_v$ is equal in magnitude and
opposite in sign to the
lepton number generated in the visible sector:
$(B-L)_{\rm v} = -L_{\rm v} \approx -0.4\kappa \epsilon s / g_*$.
where $s$ is the entropy, $\kappa$ is the washout factor and we take $\kappa = 0.1$ and $g_* = 106.75$.
Using again $B_{\rm final} \approx 0.31 (B-L)_{\rm v}$, one estimates $| \epsilon | \sim 5 \times 10^{-6}$.
The supersymmetric extension of this model is straightforward, as discussed in~\cite{Feng:2013zda}.

\subsection{Phenomenology of the model}

In a manner similar to what was discussed earlier,
the symmetric component of dark matter would be sufficiently depleted
by annihilating via the $Z'$ gauge boson into SM particles (or annihilating into $U(1)_x$ dark photons),
which ensures the asymmetric dark matter to be the dominant component of the current dark matter relic abundance.

An interesting implication of this model class arises in the neutrino sector.
Here we add three families of right-handed neutrinos.
We assume the coupling $\beta$ is family-dependent,
i.e., $\beta \to \beta_i$ where $i=1,2,3$ correspond to $e,\mu,\tau$, c.f., \cref{1.1}
so the Lagrangian reads
\begin{equation}
\mathcal{L}' =
\beta_i \bar \psi_R L_i H
+  \beta_{ij}'' \bar \nu_{iR} L_j H
+ \mu_i' \bar \nu_{iR} \psi_L
+ h.c.\,.
\end{equation}
After spontaneous breaking of the electroweak symmetry, the mass terms take
the form $\mathcal{L}_m = \vec{\nu}_R^T\,\mathcal{M} \,\vec{\nu}_L + h.c.$, where
$\vec{\nu}_R^T = \left(\bar{\nu}_{R}^{e},\bar{\nu}_{R}^{\mu},\bar{\nu}_{R}^{\tau},\bar{\psi}_{R}\right)$,
and $\vec{\nu}_L^T = \left(\nu_{L}^{e},\nu_{L}^{\mu},\nu_{L}^{\tau},\psi_{L}\right)$.
For simplicity we assume a symmetrical
form for the neutrino mass terms so that
\begin{equation}
\mathcal{L}_{m}^\nu=
\vec{\nu}_R^T
\left(\begin{array}{cccc}
m_{\nu_{e}} & 0 & 0 & \mu_{1}\\
0 & m_{\nu_{\mu}} & 0 & \mu_{2}\\
0 & 0 & m_{\nu_{\tau}} & \mu_{3}\\
\mu_1 & \mu_2 & \mu_3 & m_{1}
\end{array}\right)
\vec{\nu}_L
+h.c.\,,
\label{neutrino-matrix2}
\end{equation}
 \cref{neutrino-matrix2} contains no direct mixings among the neutrino flavor states.
However,  their mixings with the field $\psi$ automatically leads  to neutrino flavor mixings
for the mass diagonal states.
To exhibit this mixing
 we  diagonalize the matrix of \cref{neutrino-matrix2} by an orthogonal transformation.
By setting   $m_{\nu_{e}}=10^{-11},m_{\nu_{\mu}}=1.7\times10^{-10},m_{\nu_{\tau}}=2\times10^{-9},m_1=2000,
\mu_1=3.6\times10^{-5},\mu_2=8.9\times10^{-5},\mu_3=5.9\times10^{-4}$
(all masses in GeV) the three neutrino masses in the mass diagonal basis are
$m_{3} \approx 4.8 \times10^{-2}~{\rm eV},
m_{2} \approx 1.2 \times10^{-2}~{\rm eV},
m_{1} \approx 4.2 \times10^{-3}~{\rm eV}$,
which  is the normal hierarchy of neutrino masses~\cite{Beringer:1900zz}
while  the mass of  the heavy
field $\psi$ is still  $\sim m_{1}$.
For the neutrino mixings we obtain $\sin^2\theta_{12} \approx 0.30\,, \ \sin^2\theta_{23} \approx 0.36\,,\ \sin^2\theta_{13} \approx 0.024$,
which is in good accord with the experimental determination of the mixing angles.
Specifically the model is consistent with the result
from the Daya Bay reactor neutrino experiment~\cite{An:2013zwz} of $\theta_{13} \sim 9^\circ$.
It is interesting that the model provides an explanation
of the neutrino mixings at a fundamental level.
The neutrino mixings arise as a consequence
of the interaction of the neutrinos with the primordial Dirac field $\psi$ which enters in leptogenesis
which points to the cosmological origin of neutrino mixings.

\begin{figure}[t!]
  \begin{center}
  \includegraphics[scale=0.5]{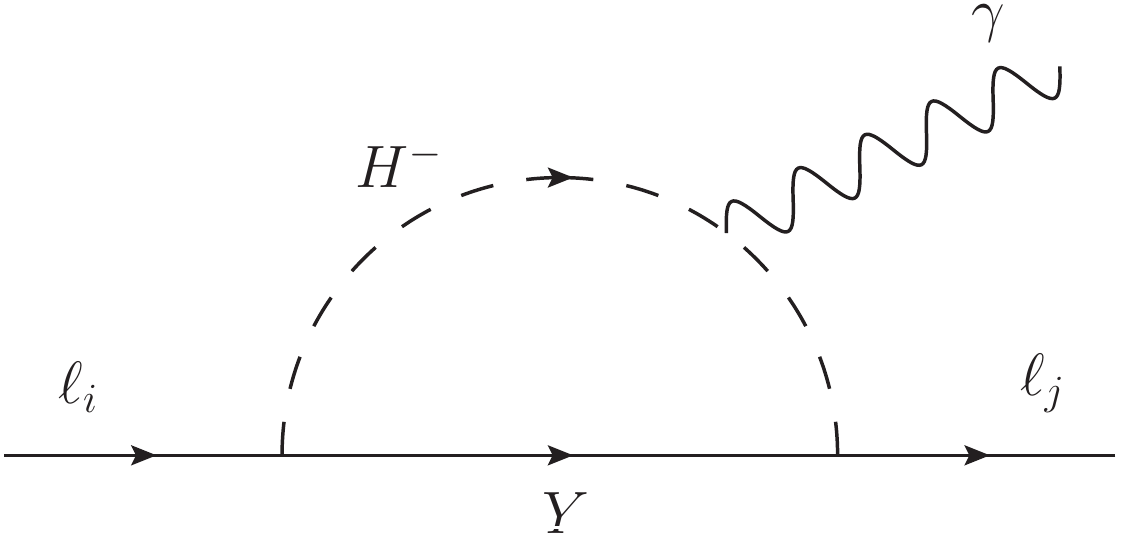}
  \vspace*{8pt}
  \caption{Flavor changing processes $\ell_i \to \ell_j \gamma$ via the charged Higgs and $Y$ loop.}
  \label{muega}
  \end{center}
\end{figure}

Other implications of the model involve flavor changing processes.
For the supersymmetric version of the model,
after spontaneous breaking one has interactions of the
charged Higgs $H^+$ with charged leptons and $Y$:
\begin{equation}
 {\cal L}_{H\ell\psi}= \beta_i \bar Y \ell_i H^+ +  h.c.\,,
\end{equation}
where $\ell_i$ denotes the charged leptons and $Y$
 is a chiral field with lepton number $-1$~\cite{Feng:2013zda}.
Such interactions will give rise to $\ell_i \to \ell_j \gamma$ processes,
where a charged lepton $\ell_i$ converts into a charged lepton $\ell_j$ via exchange of $Y$ while a photon
is emitted by the charged Higgs inside the loop,  see \cref{muega}.
Assuming $m_Y^2 \gg m_{H^+}^2$,
we obtain the decay rate of the flavor changing process $\ell_i \to \ell_j \gamma$ to be
\begin{equation}
\d \Gamma_{\ell_i \to \ell_j \gamma} = \frac{\alpha_{\rm em} (\beta_i \beta_j)^2}{(16 \pi^2)^2} \frac{m_i^3}{M_Y^2}\,,
\label{DW}
\end{equation}
where $m_i$ is the mass of the decaying charged lepton and we have used $m_i \gg m_j$.
The current experimental bounds constrain the couplings to be
$\beta_1 \sim \beta_2 \lesssim 3\times 10^{-3}$ and $\beta_3 \lesssim 2 \times 10^{-4} / \beta_1$
for $M_Y \sim 1~{\rm TeV}$.
One can expect observable effects in these flavor changing processes
in future experiments with improved sensitivities.

\section{Conclusion}

The comparable size of dark matter and visible matter in the Universe points to a possible common origin of the two.
Here we discussed two classes of models.
In the first model class, the dark matter is generated from the decay of some primordial fields.
The asymmetry is generated in the dark sector by the CP violating decays,
and then transfer to the visible sector via the asymmetry transfer interaction.
In the second model class all of the fundamental interactions conserve lepton number,
and leptogenesis occurs when equal and opposite lepton numbers are generated in the visible sector and dark sector.
Subsequently the sphaleron processes transmute a part of lepton asymmetry to baryon asymmetry.
In this model class the total $B-L$ number in the Universe is exactly conserved.
Phenomenological aspects of these models were also discussed.

\section*{Acknowledgments}
Research was supported in part by NSF Grants PHY-1314774 and PHY-1620575.

\end{document}